\documentclass{Rinton-P9x6}

\usepackage{latexsym}

\begin{document}

\title{Cosmological uses of Casimir energy}

\author{Emilio Elizalde}

\address{Department of Mathematics,
MIT, Cambridge, MA 02139
\\ E-mail: elizalde@math.mit.edu,
elizalde@ieec.fcr.es\footnote{On leave from ICE, Consejo
Superior de Investigaciones Cient\'{\i}ficas,  and IEEC, Edifici
Nexus, Gran Capit\`{a} 2-4, 08034 Barcelona, Spain.}}


\maketitle

\abstracts{A precise zeta-function calculation shows that the
contribution of the vacuum energy to the observed value of the
cosmological constant can possibly have the desired order of
magnitude albeit the sign strongly depends on the topology of the
universe. The non-renormalizable, infinite contributions which
have been recently shown to occur when one physically imposes
boundary conditions on quantum fields (Casimir calculations) are
considered. It is shown that using a Hadamard regularization in
addition to the zeta method, the ordinary, finite results in the
literature are exactly recovered.}

This report is divided into two parts. In the first one, within a
simple cosmological model, the vacuum
energy density of a scalar field with a very low (but non-zero) mass
 is seen to contribute with the
right order of magnitude to the observed value of the cosmological
constant. In the second, we elaborate on some issues, discussed
recently, concerning the  physical (and mathematical) meaning of
imposing boundary conditions (BC) on quantum fields, which is
central to Casimir effect calculations.

\section{Vacuum energy in a simple cosmological model}
It is not difficult to get the
right order of magnitude of the vacuum energy density $\rho_V$, in
the range corresponding to astrophysical observations
\cite{perl,ries}, e.g. $\rho_V \sim 10^{-10}$ erg/cm$^3$.
For this, one just assumes the existence of a scalar field background,
$\phi$, extending through the universe and calculates the
contribution to the cosmological constant coming from the Casimir
energy density \cite{Casimir} corresponding to this field for some
typical boundary conditions. (The ultraviolet contributions are
safely set to zero by invoking some mechanism of a fundamental theory.)
One further assumes
existence of both large and small dimensions (the total number of
large spatial coordinates will be always three), some of which
(from each class) may be compactified. Indeed, the global topology
of the universe plays here an important role \cite{nat03a,ct1}.
 The issue of the possible contribution of the Casimir
effect as a source of some sort of cosmic energy, as in the case
of the creation of a neutron star \cite{sokol1}, has been
considered before \cite{eejmp12,banks1,pr1,mil2,esp1}, but the
emphasis will be here put in obtaining the right order of
magnitude for the effect, with a minimum number of assumptions.

Consider a universe with a space-time as:
$\mathbf{R^{d+1}} \times \mathbf{T}^p\times \mathbf{T}^q$, or
$\mathbf{R^{d+1}} \times \mathbf{T}^p\times \mathbf{S}^q, \ldots$. A
(nowadays) free scalar field pervading the universe will satisfy
$ (-\Box +M^2) \phi =0$, restricted
by the appropriate boundary conditions (e.g., periodic, in the
first case considered).  Here, $d\geq 0$ stands for a possible
number of non-compactified dimensions.
The vacuum energy density for a ($p,q$)-toroidal universe (with
$p$  `large' and $q$  `small' dimensions) is
\begin{eqnarray}  \rho_\phi
&=&\frac{\pi^{-d/2}}{2^d\Gamma (d/2) \prod_{j=1}^p a_j \prod_{h=1}^q
b_h} \int_0^\infty dk \, k^{d-1} \sum_{\mbox{\bf
n}_p=-\mathbf{\infty}}^{\mathbf{\infty}} \sum_{\mbox{\bf
m}_q=-\mathbf{\infty}}^{\mathbf{\infty}} \nonumber \\
&& \hspace*{25mm} \left[ \sum_{j=1}^p
\left( \frac{2\pi n_j}{a_j}\right)^2 + \sum_{h=1}^q\left(
\frac{2\pi m_h}{b_h}\right)^2 +M^2 \right]^{1/2}. \label{t1}
\end{eqnarray}
We shall use zeta regularization \cite{zb1,bp1,zb3}. For the
analytic continuation of the zeta function corresponding to
(\ref{t1}) we obtain (to simplify, we consider now that all
$a$'s are equal, and also all $b$'s)  \cite{eecmp1}: \begin{eqnarray}
&&\hspace{-5mm} \zeta(s)=\frac{2\pi^{s/2+1}}{a^{p-(s+1)/2}b^{q-(s-1)/2} \Gamma
(s/2)} \sum_{\mbox{\bf m}_q=-\mathbf{\infty}}^{\mathbf{\infty}}
\sum_{h=0}^p \left(_{\,\displaystyle h\,}^{\,\displaystyle
p\,}\right) 2^h  \nonumber \\ &&\hspace{-4mm}
 \sum_{\mbox{\bf n}_h=1}^{\mathbf{\infty}}
 \left( \frac{\sum_{j=1}^h n_j^2 }{\sum_{k=1}^q
m_k^2+M^2}\right)^{(s-1)/4}\hspace{-5mm}  K_{(s-1)/2}
\left[ \frac{2\pi a}{b} \sqrt{\sum_{j=1}^h n_j^2
\left(\sum_{k=1}^q m_k^2+M^2\right)}\right], \label{z11}
\end{eqnarray} with $K_\nu (z)$  the modified Bessel function
of the second kind. Having performed already the analytic
continuation, this expression is ready for the substitution
$s=-1$, and using also the
behaviour of the function $K_\nu (z)$ for small values of its
argument, $ K_\nu (z) \sim \frac{1}{2} \Gamma (\nu)
(z/2)^{-\nu}, \, z \to 0$, we obtain, in the case when $M$ is very small,
\begin{eqnarray} \rho_\phi &=& -\frac{1}{a^pb^{q+1}} \left\{ M\, K_1 \left(
\frac{2\pi a}{b} M \right)+ \sum_{h=0}^p \left(_{ \, \displaystyle
h\,}^{\, \displaystyle p\,}\right) 2^h \sum_{\mbox{\bf
n}_h=1}^{\mathbf{\infty}} \frac{M}{\sqrt{\sum_{j=1}^h n_j^2 }}
\right.\nonumber \\ && \hspace{-3mm} \left. \times K_1\left(
\frac{2\pi a}{b} M \sqrt{\sum_{j=1}^h n_j^2} \right) + {\cal O}
\left[ q\sqrt{1+M^2} K_1\left( \frac{2\pi a}{b}\sqrt{1+M^2}\right)
\right]\right\}. \label{ff0}\end{eqnarray} We should view all
masses appearing here as dimensionless: quotients by the
mass-dimensional regularization parameter $\mu$, as $M/\mu$
everywhere. This does not affect the small-$M$ limit, which reads
$M/\mu << b/a$. Replacing in the expression the $\hbar$ and $c$
factors, we get
\begin{eqnarray} \rho_\phi = -\frac{\hbar c}{2\pi a^{p+1}b^q}
\left[1+\sum_{h=0}^p \left(_{\,\displaystyle h\,}^{\,\displaystyle
p\,}\right) 2^h \alpha \right]+ {\cal O} \left[ q K_1\left(
\frac{2\pi a}{b}\right) \right], \label{ff1}
\end{eqnarray} with $\alpha$ a finite constant (an explicit
geometrical sum in the limit $M\rightarrow 0$).

For the most common variants, the constant $\alpha$ in (\ref{ff1})
has been calculated to be of order $10^2$, and the whole factor,
in brackets, of the first term in (\ref{ff1}) is of order $10^7$.
Good coincidence with the observational value for the cosmological
constant is obtained for the contribution of a very low mass ($M
\leq 1.2 \times 10^{-32}$ eV) scalar field, $\rho_\phi$, for
$p=0,1$ large compactified dimensions ($a$ the radius of the
observable universe) and $q=p+1$ small compactified dimensions,
the small compactification length, $b$, being of the order of 100
to 1000 times the Planck length $l_P$. The best fit is obtained
for $p=1, q=2$, which is  a very reasonable result, coinciding
with other much more fundamental approaches. Dimensionally
speaking, everything is here dictated by the two basic lengths in
the problem: its Planck value and the radius of the observable
Universe, and the final conclusion appears to be quite robust.

\section{Hadamard regularization of the Casimir effect}

We luckily chose in the previous section a no-boundary
configuration so as to avoid the important problem we will now address:
that of imposing BC on a quantum field \cite{dcs}.
To go directly to the heart of the matter, take again the most 
simple case of  a scalar field in one dimension,
$\phi (x)$, with a BC of Dirichlet type imposed at a point, e.g.
$\phi (0)=0$. We want to calculate the Casimir\cite{Casimir}
energy for this configuration, that is, the difference between the zero point
energy  corresponding to such field when the BC is enforced, and
the zero point energy in the absence of any BC. Both energies are
infinite and the regularized difference may
still be infinite when the BC point is approached (this is the
result in \cite{bj1}) or may be finite (even zero,
which is the result given in many standard references on the subject
\cite{cb1,mil3}).

\subsection{Understanding the infinities  `ab initio'}

Let us try to understand this enormous discrepancy \cite{een}. To this 
end, we propose to go back to the more classical (and mathematical) definition
of boundary value problem, as taken e.g.~from Courant and Hilbert (see 
Ref.~\cite{ch1}, Chap.~V, pp.~275 ff). Unfortunately, there is here no place 
to go into details. Suffice to say that, within this definition, 
 one imposes smoothness of the solution on the points of the 
boundary itself (which are not different, in this way, from the rest 
of the points).\footnote{For a number of
 physical applications, as reflection of wave packets, BC for
perfectly conducting walls, or bag type conditions, it would not
be adequate to demand this: the boundary is explicitly excluded from the
domain, which is usually broken by it into two or more independent 
subdomains.} Thus, the boundary is not supposed to divide the
space into independent domains (or`universes', so to speak). It
can be argued that this sort of mathematical boundary value
problem is better suited for the analysis of the quantum
vacuum \cite{een}, in view that the other definition cannot solve the
dilemma posed by Bob Jaffe and collaborators. Indeed, in the
papers \cite{bj1}, and in Jaffe's talk at this workshop, it became
cristal clear that the ordinary definition\footnote{The one not demanding
regularity on the boundary.} is quite useless for describing
the Casimir energy density: it leads to misleading results which
have nothing to do with the physics of the Casimir effect in QFT 
 \cite{bj1}.

And here comes the calculation itself. One has to add up all energy
modes (trace of $H$). For the mode with energy $\omega$, the field
equation is:
\begin{eqnarray}
-\phi''(x)+m^{2}\phi (x)=\omega^{2}\phi(x). \label{fe1}
\end{eqnarray}
In the absence of a BC, the solutions to the field equation can be
labelled by $k = + \sqrt{\omega^2 - m^2} > 0$, as
$\phi_k (x)= A e^{ikx} + B e^{-ikx}$,
with $A,B$ arbitrary complex (for the general complex), or as
$\phi_k (x)= a \sin (kx) + b \cos (kx)$,
with $a,b$ arbitrary real (for the general real solution). Now,
when the mathematical BC of Dirichlet type, $\phi (0)=0$, is
imposed, this does not influence at all the eigenvalues, $k$,
which remain exactly the same.
However, the {\it number} of solutions corresponding to each
eigenvalue is reduced by one half to:
$\phi_k^{(D)} (x)= A (e^{ikx} -  e^{-ikx})$,
with $A$ arbitrary complex (complex solution), and $\phi_k^{(D)}
(x)= a \sin (kx)$, with $a$ arbitrary real (real solution). In
other words, in both cases is the same, a continuous spectrum, but
the number of eigenstates corresponding to a given eigenvalue is
twice as big in the absence of the BC \cite{ch1}.

An infinity may originate from the fact that imposing the BC has
drastically reduced to one-half the family of eigenfunctions for
the spectrum of the operator. And, since this dramatic reduction
takes place precisely at the point where the BC is imposed, a
physical divergence (infinite energy) may originate there. While
this sketchy analysis cannot be taken as a substitute for the
actual modelization of Jaffe et al. \cite{bj1}
---where the BC is explicitly enforced through the introduction of
an auxiliary, localized field, which probes what happens at the
boundary in a much more precise way--- it certainly leads one to
think that pure mathematical considerations, which include the use
of analytic continuation by means of the zeta function, are in no
way blind to the infinites of the physical model and do not
produce misleading results, when the mathematics are used
properly. And it is very remarkable to realize how close the
mathematical description of the appearance of an infinite
contribution is to the one provided by the more physical
realization [see R. Jaffe et al's contributions to these
proceedings].

What I have shown with this little exercise is precisely that,
e.g., by defining the 1-dimensional Dirichlet boundary value 
problem in a classical (or more mathematically minded) fashion, 
which views the solution of the eigenvalue problem on
the whole real line ---before and after forcing it to be zero at
the origin--- we do gain a mathematical understanding of the
emergence of a singularity at the boundary, which very
closely parallels the physical description of Jaffe et al.
\cite{bj1} This is certainly not the final answer, since the
divergence should be now regularized and renormalized,  and the
whole argument is based on a maybe non-standard\footnote{Better,
not-so-useful-in-other-cases.} definition of boundary value
problem, but it works in this situation (and not only for one
dimension \cite{een}). The reason why these infinities do not
usually show up in the literature on the Casimir effect \cite{cb1}
is probably because the other definition of boundary value problem
was used generically,\footnote{The one which does {\it not} impose 
regularity of the
eigensolution at the boundary.} and also because textbooks on the
subject often focus towards the calculation of the Casimir  force
(minus the derivative of the energy). Since the infinite terms do
not depend on the distance between the plates, they do not
contribute to the force (see also \cite{mil3}).

\subsection{Splitting the infinities}

In the rest of the paper we will focuss on the interpretation of 
the persistent infinities obtained  in the QFT calculations, i.e.
those which remain after renormalization \cite{bj1}. Surprisingly, 
by using
Hadamard regularization we are going to see that one can split
those divergences, in a very natural way, into two terms: a finite 
and an infinite contribution, each of them having a very precise 
interpretation.
Indeed, the finite part will be recognized as the finite result
obtained in the already mentioned usual references on the Casimir 
effect. The remaining infinite part will be precisely identified 
as proportional to $1/\omega$, where $\omega$ is typically the  
width of the Gaussian that gives rise to the delta-function 
(when $\omega \to 0$),
which enforces the BC in the approach of \cite{bj1}.
This is undoubtedly a remarkable result.

The naturalness of the splitting of the divergences just mentioned 
hangs on that of Hadamard's regularization itself, as a standard and 
admissible technique. Now, Hadamard regularization is actually a 
well-established procedure in order
to give sense to infinite integrals: in higher-post-Newtonian
general relativity \cite{had1} and also in
 QFT \cite{had2}. Among mathematicians,
it is the standard technique in order to deal with singular
differential and integral equations with BCs, both analytically
and numerically. 

In one dimension, with Dirichlet BC imposed at
one ($x=0$) and two ($x=\pm a$) points, respectively, by means of
a delta-background of strength $\lambda$ (see \cite{bj1}), one
encounters the two divergent integrals:
\begin{eqnarray}
&& E_{1}(\lambda,m)  = \frac{1}{2\pi}\int_{m}^{\infty}
\frac{dt}{\sqrt{t^2-m^2}}\,
\left[t\log\left(1+\frac{\lambda}{2t}\right)-\frac{\lambda}{2}\right],
 \label{d11} \\
&& \hspace{-6mm} E_2(a,\lambda,m) = \frac1{2\pi} \int_m^\infty
\frac{dt}{\sqrt{t^2-m^2}}\,
           \left\{ t\log \left[1+\frac\lambda{t} + \frac{\lambda^2}{4t^2}
           \left(1-e^{-4at}\right)\right] - \lambda\right\}.
           \label{d12}
\end{eqnarray}
Using Hadamard's regularization, as described before, we get
for the first one
\begin{eqnarray}
 E_{1}(m) = \left. \frac{\lambda}{4\pi} \left(1-\ln \frac{\lambda}{m} \right)
\right|_{\lambda \to \infty}  + \
 =\hspace*{-4mm}\int \ \, , \label{e11}
\end{eqnarray}
where the first term is the singular part when the limit $\lambda
\to \infty$ is taken, and the second ---which is Hadamard's finite
part--- yields in this case
\begin{eqnarray}
  =\hspace*{-4mm}\int \ = \, -\frac{m}{4}.
\end{eqnarray}
Such result is coinciding with  the classical one (0, for $m=0$).
Note in particular, that the further $ \ln m $ divergence as $ m
\to \infty $ is hidden in the  $\lambda-$divergent part, and such
behavior does explain why the classical results\cite{cb1} which
are obtained using hard Dirichlet BC ---what corresponds as we
just prove here to the Hadamard's regularized part--- cannot see
it \cite{een}.

In the case of a two-point boundary at $x=\pm a$ (separation
$2a$), Eq.~(\ref{d12}), we get a similar Eq.~(\ref{e11}) but now
the regularized integral is as follows. For the massless case, we
obtain
\begin{eqnarray}
  =\hspace*{-4mm}\int \ = \, -\frac{\pi}{48 a},
\end{eqnarray}
which is  the `classical' regularized result. In the massive case,
$m\neq 0$, after additional work the following fast convergent
series turns up
\begin{eqnarray}
 =\hspace*{-4mm}\int \ = \,  -\frac{m}{2\pi} \sum_{k=1}^\infty
\frac{1}{k} K_1 (4akm).
\end{eqnarray}
Thus Eq.~(\ref{e11}) yields strictly the same result which one
already obtains by imposing the Dirichlet BC {\it ab initio} in the
ordinary way \cite{een}. What has now been {\it gained} is a more clear
identification of the singular part, in terms of the strength of
the delta potential at the boundary. That is the general
conclusion of this analysis, common to all the other cases here
considered. (For an alternative approach which also uses Hadamard
calculus see the contribution of S. Fulling to these proceedings.)

Correspondingly, for the Casimir force we obtain the finite values
\begin{eqnarray}
 F_2(a) = -\frac{\pi}{96 a^2},
\end{eqnarray}
in the massless case, and in the massive one
\begin{eqnarray}
  F_2(a,m)  =  -\frac{m^2}{\pi} \sum_{k=1}^\infty \left[
K_0 (4akm) +\frac{1}{4akm} K_1 (4akm) \right].
\end{eqnarray}
Those expressions coincide with the ones derived in the above
mentioned textbooks on the Casimir effect, and reproduced before
by using the zeta function method (for a recent, very simple 
derivation see \cite{et1}).

The two dimensional case turns out to be more singular. After a
similar analysis, by using Hadamard regularization we also obtain
the usual finite classical results of the literature in this case.
For the Casimir energy, we get
(notation as in \cite{bj1})
\begin{eqnarray}
&&\hspace*{-8mm} E^{(2)}_{\lambda^2} [\tau] =
\left. \frac{\lambda^2a^2}{8}\int_0^\infty dp \,
(ap+1)^{-\omega} J_0(ap)^2 \arctan (p/2m)  \right|_{\omega \to 0} \nonumber \\
&&\hspace*{-3mm} = \frac{\lambda^2a^2}{8} \left\{
 \frac{1}{2\omega} \ +  \frac{\gamma + 3 \ln 2}{2a}
+4m \left[ \gamma - \frac{2}{\sqrt{\pi}} \left[ 1- \ln (am)
\right] \, h(4a^2m^2) \right]\right\},
\end{eqnarray}
where $ h(z):= {_2F_3} \left( (1/2,1/2);(1,1,3/2);z \right)$ and
$\gamma$ is the Euler-Mascheroni constant; in particular,
 $h(1)=1.186711$, quite a nice value. Here, as previously announced, 
  $ \omega$ is the  width of the Gaussian
 $ \delta$, which is
the very  physical parameter considered by Candelas \cite{dcs}. The
finite part (Hadamard) is
\begin{eqnarray}
=\hspace*{-4mm}\int_0^\infty   dp\, J_0(ap)^2 \arctan (p/2m).
\end{eqnarray}
Again, this reverts to the results obtained in the
literature using Dirichlet BC {\it ab initio}.

\section*{Acknowledgments}
I am indebted to the members of the Mathematics
Department, MIT, and specially to Dan Freedman, for warm
hospitality. I thank the anonymous referee of this contributed 
paper for critical insights and patient discussions which definitely 
led to its improvement. 
The present investigation has been supported by DGICYT
(Spain), project BFM2000-0810, and by CIRIT (Generalitat de
Catalunya), grants 2002BEAI400019 and 2001SGR-00427.

\end{document}